%% file: main.tex
\documentclass[10pt,twocolumn,prb,aps,amsmath,amssymb,superscriptaddress]{revtex4-2}

\usepackage{graphicx}
\usepackage{dcolumn}
\usepackage{bm}
\usepackage{float}
\usepackage[colorlinks=true,linkcolor=blue,citecolor=blue,urlcolor=blue]{hyperref}
\usepackage{natbib}
\usepackage{color}
\usepackage[normalem]{ulem}

\begin{document}

\title{
A Hybrid Brownian Dynamics–kinetic Monte Carlo Framework for Coupled Magnetic and Rotational Dynamics of Nanoparticles
}

\author{Siraj Ul Haq}
\email{siraj.ulhaq@usc.es}
\affiliation{Departamento de Física Aplicada, Universidade de Santiago de Compostela, 15782 Santiago de Compostela, Galicia, Spain}
\affiliation{Instituto de Materiais (iMATUS), Universidade de Santiago de Compostela, 15782 Santiago de Compostela, Galicia, Spain}

\author{Iago López-Vázquez}
\affiliation{Departamento de Física Aplicada, Universidade de Santiago de Compostela, 15782 Santiago de Compostela, Galicia, Spain}
\affiliation{Instituto de Materiais (iMATUS), Universidade de Santiago de Compostela, 15782 Santiago de Compostela, Galicia, Spain}

\author{Kazuya Okada}
\affiliation{Department of Mechanical Engineering, Saitama Institute of Technology, Fukaya, Japan}

\author{Sergiu Ruta}
\affiliation{College of Business, Technology and Engineering, Sheffield Hallam University, UK}

\author{Amira P\'aez-Rodr\'iguez}
\affiliation{Instituto de Nanociencia y Materiales de Aragón (INMA) CSIC-Universidad de Zaragoza  and CIBER-BBN, Spain.}

\author{Luc\'ia Guti\'errez}
\affiliation{Instituto de Nanociencia y Materiales de Aragón (INMA) CSIC-Universidad de Zaragoza  and CIBER-BBN, Spain.}

\author{Òscar Iglesias}
\affiliation{Departamento de Física de la Matèria Condensada, Universitat de Barcelona and IN2UB, c/ Mart\'{\i} i Franquès 1, 08028 Barcelona, Catalunya, Spain}

\author{David Serantes}
\affiliation{Departamento de Física Aplicada, Universidade de Santiago de Compostela, 15782 Santiago de Compostela, Galicia, Spain}
\affiliation{Instituto de Materiais (iMATUS), Universidade de Santiago de Compostela, 15782 Santiago de Compostela, Galicia, Spain}

\author{Roy W. Chantrell}
\email{roy.chantrell@york.ac.uk}
\affiliation{School of Physics, Electronics and Technology, The University of York, York, UK}



\begin{abstract}
Simulating the coupled magnetic and rotational dynamics of suspended magnetic nanoparticles (MNPs) over experimentally relevant timescales remains computationally demanding, particularly under low-frequency alternating magnetic fields. Here, we develop a hybrid Brownian Dynamics--kinetic Monte Carlo (BD--kMC) framework that combines thermally activated magnetic switching with particle rotation while retaining local fluctuations of the magnetic moment. These fluctuations are incorporated through Metropolis Monte Carlo (MC) sampling, allowing the moment to depart from the energy minima and thereby capturing the thermally averaged anisotropy torque that governs particle rotation. To preserve the distinction between local fluctuations and magnetic reversals, a barrier-trapping criterion confines MC trial moves to the current energy well, leaving thermally activated switching to the kMC process. For non-interacting single-domain MNPs, comparison with stochastic Landau--Lifshitz--Gilbert (LLG) simulations demonstrates agreement in the equilibrium torque and stationary easy-axis orientation, including the field-driven transition from predominantly perpendicular to parallel alignment. The resulting BD--kMC framework provides an efficient approach to coupled magnetic and Brownian Dynamics, reducing the computational cost of exploring the long timescales relevant to low-frequency magnetic applications.

\end{abstract}

\maketitle

\section*{Keywords}
Magnetic nanoparticles; Brownian Dynamics; kinetic Monte Carlo; Easy-axis reorientation; Anisotropy torque; Barrier-trapping

\section{Introduction}

Magnetic nanoparticles (MNPs) have attracted significant interest owing to their magnetic response to static and alternating magnetic fields, enabling a wide range of biomedical applications, including magnetic nanoparticle hyperthermia (MNH)~\cite{Rosensweig2002,Laurent2011,Deatsch2014}, targeted drug delivery~\cite{Pankhurst2003,Neuberger2005}, and magnetic particle imaging (MPI) for real-time diagnostic imaging~\cite{Gleich2005}. Their performance is governed by two fundamental magnetic relaxation mechanisms~\cite{Pankhurst2003,Deatsch2014}: Néel relaxation, in which the magnetic moment reorients within the particle by crossing an anisotropy energy barrier~\cite{Brown1963}, and Brownian relaxation, where the particle itself physically rotates within its host medium~\cite{Ilg2020}. The relative contribution of these mechanisms depends on particle size, magnetic anisotropy, fluid viscosity, and field amplitude and frequency, making their accurate description essential for optimizing MNP-based applications. 

To understand the magnetic response, it is important to consider the rotational dynamics of the particle easy axis during Brownian motion. Under an AC field, the magnetic moment evolves continuously through the combined effects of thermal fluctuations, magnetic anisotropy, and external magnetic field, while the particle itself undergoes Brownian motion within the surrounding fluid. The resulting reorientation of the particle easy axis strongly influences the magnetic relaxation and energy dissipation. Although the external magnetic field exerts a magnetic torque on the magnetic moment, this torque alone does not account for the coupling between magnetization and particle easy axis~\cite{okada2025proposal, wolfschwenger2024molecular}. Instead, the anisotropy torque directly governs this coupling and provides a consistent description of particle rotation during the coupled Brownian and Néel relaxation processes~\cite{usov2012dynamics}. Therefore, accurately evaluating the anisotropy torque is essential for predicting the easy-axis orientation under an external magnetic field.  

Several studies have highlighted the importance of easy-axis dynamics in determining the magnetic response of nanoparticles under an AC field. Shah \textit{et al.}~\cite{Shah2015} demonstrated that the Brownian alignment of particle easy axes can significantly affect the magnetic response, even when Néel relaxation dominates the magnetization dynamics.
Ota \textit{et al.}~\cite{Ota2019} investigated the effects of particle size and magnetic anisotropy on easy-axis dynamics and the resulting harmonic signals in MPI.
Their subsequent work further demonstrated that the rotation and alignment of the easy axis depend on Brownian relaxation, magnetic anisotropy, and the applied field~\cite{Ota2021}. In addition, Shasha and Krishnan~\cite{Shasha2021} showed that the coupled effects of Brownian rotation, Néel relaxation, magnetic anisotropy, and time-dependent magnetic fields are essential for accurately describing the non-equilibrium magnetization dynamics of MNPs. Despite these advances, accurately modeling easy-axis reorientation over a wide range of field amplitudes and frequencies remains challenging.

To address this, several analytical and numerical approaches have been developed. Analytical approaches, such as Rosensweig's theory of magnetic hyperthermia~\cite{Rosensweig2002} and the dynamic susceptibility model developed by Shliomis~\cite{Shliomis1994}, provide valuable insight into Brownian and N\'eel relaxation mechanisms. However, these models rely on simplifying assumptions--including equilibrium conditions, linear response, and idealized particle shapes--which limit their applicability to systems exhibiting strong anisotropy and non-linear magnetic behavior. Numerical methods overcome these limitations by explicitly resolving the stochastic evolution of the magnetic moment. 
One of the standard methods for simulating particle dynamics is the Landau–Lifshitz–Gilbert (LLG) equation, which resolves the precessional, damped and thermally driven evolution of the magnetic moment~\cite{Berkov1993,GarciaPalacios1998}. However, this becomes computationally expensive for long simulation timescales~\cite{okada2025proposal,wolfschwenger2024molecular}. Some LLG-based studies have focused on restricted physical conditions; for example, de Ch{\'a}tel \textit{et al.} assumed that the particles were fixed in position and orientation and analyzed only Néel relaxation~\cite{deChatel2009}. 

The conventional Monte Carlo (MC) method efficiently samples the equilibrium configurations~\cite{Metropolis1953}; however, it lacks direct correspondence with physical time. Kinetic Monte Carlo (kMC) methods~\cite{Chantrell2000} and double-well rate-equation models~\cite{Nikolaos2026} overcome this limitation by introducing transition rates based on thermally activated energy barrier crossing, enabling simulations over experimentally relevant timescales. However, these approaches do not account for the local thermal fluctuations of the magnetic moment within an energy well. 

Several hybrid approaches have been proposed to combine efficient descriptions of magnetic switching with Brownian particle rotation. Ilg~\cite{Ilg2019} introduced a diffusion-jump model combining rotational Brownian Dynamics with thermally activated Néel reversals for non-interacting nanoparticles in the high anisotropy barrier regime. Ilg and Kröger~\cite{Ilg2020} subsequently extended this framework to interacting particle systems using a hybrid BD--kMC method, while Wolfschwenger \textit{et al.}~\cite{wolfschwenger2024molecular} developed a hybrid Langevin--kMC model for interacting particle ensembles. 
In both kMC descriptions, the magnetic dynamics is represented by thermally activated switching between stable states, whereas small thermal fluctuations around these states are not explicitly described. 
In contrast, Okada and Satoh~\cite{okada2025proposal} proposed a hybrid BD--MC method in which Brownian Dynamics governs particle rotation and the magnetic moment is updated using Metropolis MC sampling. Although this hybrid method significantly reduces the computational cost compared with stochastic LLG simulations, conventional Metropolis MC sampling has no direct correspondence with physical time and therefore does not directly provide the timescale of Néel switching. 

Motivated by these limitations, we propose a hybrid BD--kMC model augmented with Metropolis MC sampling, which provides a consistent description of Brownian rotation, Néel relaxation and thermal fluctuations under an AC field over experimentally relevant timescales. The MC cone angle ($\delta\theta$) and the number of MC trial steps per Brownian time step are systematically optimized to ensure accurate coupling between Brownian rotation and N\'eel relaxation. In addition, a barrier-trapping scheme is introduced to prevent unphysical barrier-crossing events during MC sampling. The performance of the proposed approach is assessed against stochastic LLG simulations over a broad range of magnetic field amplitudes and frequencies. The reference LLG dynamics are integrated using an adaptive Dormand--Prince DP45 scheme, in which the integration time step is dynamically adjusted according to the local numerical error, following the implementation described in Ref.~\cite{lopez2026ac}. This comparison demonstrates that the hybrid model accurately reproduces the easy-axis dynamics while substantially reducing the computational cost.

The remainder of the paper is organized as follows. Section~\ref{sec:theo_model} introduces the theoretical framework of the  model, including the magnetic energy, thermally activated kMC switching, Metropolis MC sampling, Brownian rotational dynamics, and the coupling between these components. 
Section~\ref{sec:Results} 
presents the validation and dynamical analysis of the model, including the equilibrium torque analysis and thermal decoupling, the validation of the MC sampling procedure against the Boltzmann distribution, the coupled magnetic and Brownian Dynamics, the analysis of switching events occurring during MC moves, and the introduction of the barrier-trapping criterion. 
Section~\ref{sec:freque} investigates the frequency dependence of the easy-axis reorientation and evaluates the computational efficiency of the hybrid BD--kMC approach relative to stochastic LLG simulations. 
Section~\ref{sec:conclusions} summarizes the main conclusions.

\section{Theoretical Model} \label{sec:theo_model}

\begin{figure}[htb!]
    \centering
    \includegraphics[width=0.5\columnwidth]{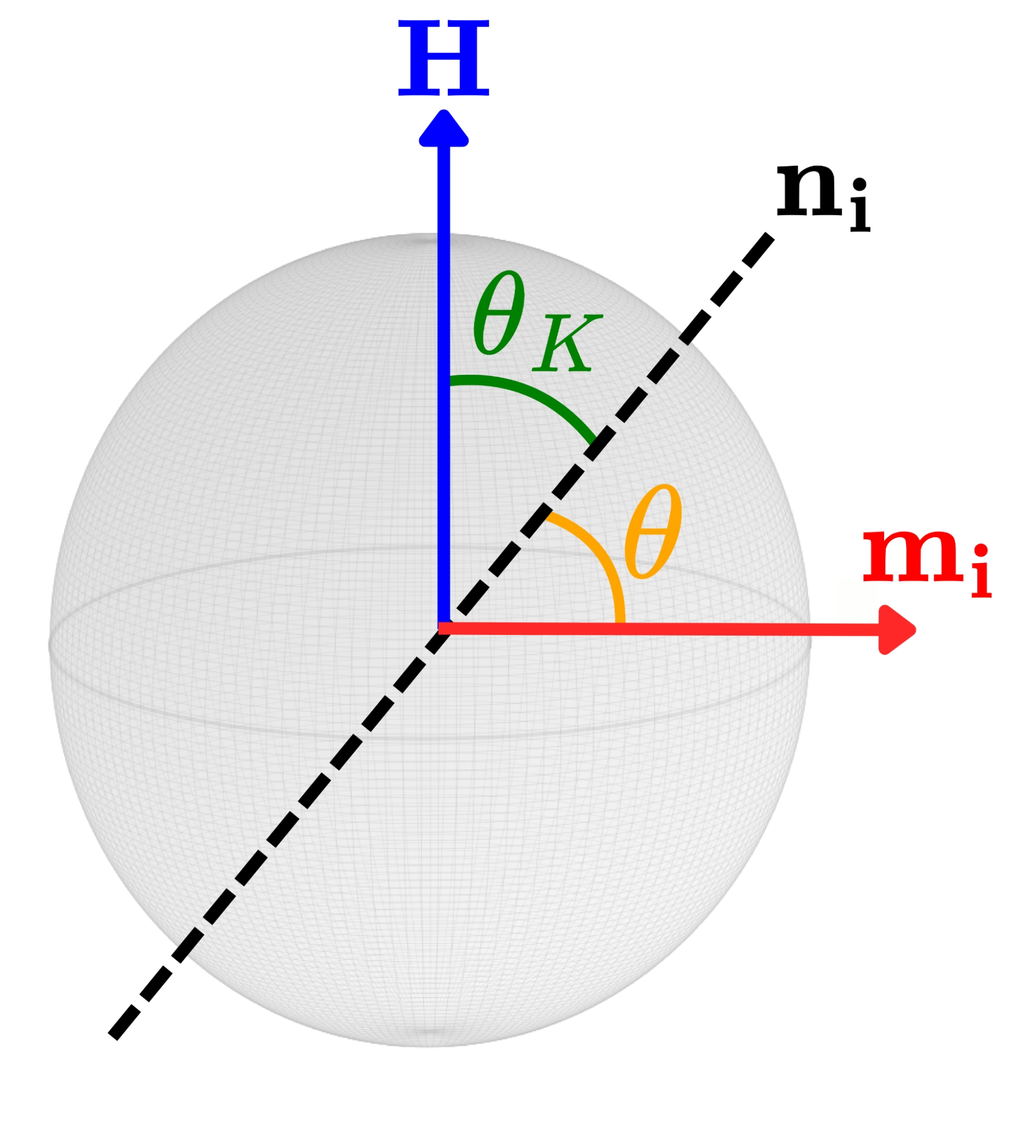}
    \caption{Schematic representation of the single-domain nanoparticle considered in the model. The unit vector \(\mathbf{m}_i\) denotes the direction of the magnetic moment, \(\mathbf{n}_i\) represents the uniaxial anisotropy easy axis, and \(\mathbf{H}\) is the externally applied magnetic field. The angle \(\theta\) is defined between \(\mathbf{m}_i\) and \(\mathbf{n}_i\), while \(\theta_K\) denotes the angle between \(\mathbf{n}_i\) and \(\mathbf{H}\).}
    \label{fig:scheme}
\end{figure}

We developed a hybrid BD--kMC framework to simulate the coupled evolution of the magnetic moments and particle orientations of  single-domain MNPs under static and alternating fields. The objective is to efficiently reproduce the long-time magnetic dynamics while retaining the essential thermal fluctuations that control the magnetization reversal and  rotational Brownian motion.  
An ensemble of 1000 non-interacting spherical MNPs with a diameter of 30 nm suspended in a fluid with a dynamic viscosity of $\eta = 0.00089~\mathrm{Pa\,s}$ at $T = 300~\mathrm{K}$ is considered. The particle positions are fixed, so
translational motion is excluded and only rotational Brownian Dynamics is considered. Each nanoparticle is treated as an independent single-domain particle, meaning that non-uniform magnetization states are not considered. The magnetic state of each particle is fully described by a unit vector $\mathbf{m}_i$, representing the direction of its magnetic moment, and an easy-axis unit vector $\mathbf{n}_i$, which defines the uniaxial anisotropy direction of the particle, as schematically shown in Fig.~\ref{fig:scheme}. 

The total magnetic energy of each nanoparticle consists of uniaxial and Zeeman contributions,
\begin{equation}
E_i = -KV(\mathbf{m}_i \cdot \mathbf{n}_i)^2 - \mu_0 M_s V\, \mathbf{m}_i \cdot \mathbf{H},
\end{equation}
where $K$ is the uniaxial anisotropy constant, $V$ is the particle volume, $\mu_0$ is the vacuum permeability, $M_s$ is the saturation magnetization, and $\mathbf{H}$ is the applied magnetic field. In the present work, magnetite ($\mathrm{Fe_3O_4}$) nanoparticles are considered, with $K = 1.1\times10^{4}~\mathrm{J\,m^{-3}}$ and $M_s = 480\times10^{3}~\mathrm{A\,m^{-1}}$. We adopt a simplified uniaxial anisotropy model; the cubic
magnetocrystalline anisotropy of magnetite is not included
and lies beyond the scope of the present study.

\subsection{kinetic Monte Carlo (kMC): Thermally Activated Magnetic Switching}

The kMC algorithm is used to model rare thermally activated transitions between the magnetic states separated by energy barriers and avoids the need for continuous integration of the LLG equation. In a single-domain nanoparticle, these barriers arise mainly from uniaxial anisotropy energy, which creates two energetically favorable orientations aligned with the easy axis. The switching rate from a given magnetic state is described by~\cite{Brown1963}:
\begin{equation}
f_i = f_0 \exp\left(-\frac{\Delta E_i}{k_B T}\right),
\label{eq:switching}
\end{equation}
where $f_0$ is the attempt frequency, $\Delta E_i$ is the energy barrier separating two magnetic states, $k_B$ is Boltzmann's constant, and $T$ is the temperature. This expression reflects the fact that thermal fluctuations enable the system to overcome energy barriers. In general, the easy axis and magnetic field are oriented at some angle $\theta_K$. The energy barrier for this case is calculated using a numerical approximation due to Pfeiffer~\cite{Pfeiffer1990}:
\begin{equation}
\Delta E(\theta_K ,\boldsymbol{H_{loc}}) = KV \left[ 1- \dfrac{| {H_{loc}|/H_k}} {g(\theta_K)}\right]^{\kappa(\theta_K)},
\label{Eq:Hubert}
\end{equation}
where $g(\theta_K)=  [ \cos^{2/3}\theta_K + \sin^{2/3}\theta_K ] ^{-3/2} $, $\kappa(\theta_K) = 0.86 + 1.14 g(\theta_K) 
$, and $H_k = 2K/(\mu_0 M_s)$ is the anisotropy field. 
Eq.~\ref{Eq:Hubert} is a numerical fit to the actual calculated energy barrier and is a remarkably good approximation.
The probability that a switching event between energy minima occurs within a finite time interval $\Delta t$ is given by:
\begin{equation}
P_i = 1 - \exp(-f_i \Delta t).
\end{equation}
Instead of explicitly integrating the fast precessional motion, the kMC method advances the system directly between the thermally activated switching events, allowing simulation over experimentally relevant timescales. 

This approach does not explicitly introduce superparamagnetic particles where the energy barriers are sufficiently small that barrier crossings in both directions are similar in magnitude. An improved approach to the computational treatment of superparamagnetic particles can be derived~\cite{Chantrell2000} by considering the relaxation of the magnetic moment described in general terms by the master equation for a two-state system. This leads to the condition that, if the transition is allowed, the moment is assigned to either minimum with a probability
\begin{equation}
p_i=e^{-E_{i}/k_BT}/(e^{-E_{1}/k_BT}+e^{-E_{2}/k_BT})
\end{equation}
with $i=1,2$  labelling the minima. Physically this process ensures that the population of the two states obeys the Boltzmann distribution in thermal equilibrium. This process applies when there is a well-defined two-state system, i.e., when the energy barrier is large enough that the moment is constrained to lie close to one or other of the energy minima. For $KV/k_B T \lesssim 3$, thermal fluctuations become comparable with the energy barrier such that kinetic Monte Carlo alone does not accurately reproduce the equilibrium state. Therefore, Metropolis MC sampling is used to ensure correct Boltzmann statistics and accurate equilibrium magnetization curves \cite{RutaMScThesis} and linear response behavior \cite{Ruta2015Hyperthermia}.
For such small energy barriers, we use a standard Metropolis MC algorithm~\cite{Metropolis1953}, as described in the following section.

\subsection{Metropolis MC: Local Magnetic Moment Fluctuations}

Although the kMC algorithm accurately reproduces thermally activated barrier crossing, it does not describe the continuous fluctuations of the moment within an energy minimum. These missing fluctuations are incorporated by Metropolis MC trial moves~\cite{Metropolis1953}.
For each trial orientation, the change in magnetic energy, 
\begin{equation}
\Delta E = E_{\mathrm{trial}} - E_{\mathrm{current}},
\end{equation}
is evaluated, and the trial move is accepted with probability
\begin{equation}
P_{\mathrm{acc}} = \min\left(1, \exp\left(-\frac{\Delta E}{k_B T}\right)\right).
\end{equation}
These trial moves preserve the Boltzmann equilibrium distribution, while allowing the evolution of the magnetic moment around the local energy minimum. It is important to note that, after each step of the kMC algorithm, MC moves are carried out to ensure that the system evolves into the required Boltzmann distribution. As shown later, this thermal decoupling between the magnetization and the easy axis is essential for correctly evaluating the torque that drives nanoparticle rotation. For each MC trial, two orthogonal unit vectors perpendicular to the current magnetic moment direction were constructed. Two independent random coefficients were generated uniformly with a maximum magnitude of $6\pi/180$. The resulting vector was then normalized to obtain the trial magnetic moment direction.

\subsection{Brownian Rotational Dynamics}

The rotational motion of each nanoparticle is described using  Brownian Dynamics~\cite{Satoh2017, wolfschwenger2024molecular}. The easy-axis orientation evolves according to 

\begin{equation}
\frac{d\mathbf{n}_i}{dt}=
\left(\frac{\mathbf{T}_i}{\gamma_r}+\boldsymbol{\xi}_i^{\,r}(t)
 \right)\times \mathbf{n}_i
,
\end{equation}
where $\gamma_r$ is the rotational friction coefficient, $\mathbf{T}_i$ is the deterministic torque acting on the particle $i$, and $\boldsymbol{\xi}_i^{\,r}(t)$ represents the random angular velocity satisfying

\begin{equation}
\langle \xi_{i\alpha}^{\,r}(t) \rangle = 0, \quad
\langle {\xi}_{i\alpha}^{\,r}(t)\,{\xi}_{j\beta}^{\,r}(t') \rangle
= 2D_r \, \delta_{ij}\delta_{\alpha\beta}\,\delta(t-t'),
\end{equation}
where $\alpha, \beta = x, y,z $, and $D_r=k_BT/\gamma_r$ is the rotational diffusion coefficient. After each BD time step, the updated easy-axis vector is normalized to preserve its unit magnitude. The deterministic torque is obtained from the anisotropy energy 
\begin{equation}
\mathbf{T}_{ani} = -2KV (\mathbf{m}_i\cdot\mathbf{n}_i)\, (\mathbf{m}_i\times\mathbf{n}_i),
\end{equation}
which couples the magnetic moment with particle orientation. 

\subsection{Hybrid BD–kMC algorithm}
The hybrid simulation combines the three numerical algorithms within each simulation cycle. First, the kMC algorithm determines whether a switching event occurs during the current time interval. If a transition is accepted, then MC steps are performed to sample the local thermal fluctuations while preventing unphysical barrier crossing.
The updated magnetic moment is then used to evaluate the anisotropy torque, which drives the Brownian rotational update of the easy-axis orientation. This hybrid coupling ensures consistent exchange of information between the magnetic switching and particle rotation throughout the simulation. It is important to note that the anisotropy torque is averaged over the Brownian Dynamics time step in order to take account of the thermal decoupling of the magnetization and easy axis referred to earlier. The hybrid scheme is illustrated in Fig. S1 of the Supplementary Information.

\section{Results and Discussion} \label{sec:Results}

We proceed with an investigation of the coupled spin and rotational dynamics using the hybrid model described above. Initially, we investigate the equilibrium torque on a nanoparticle at a fixed orientation to the applied field, which demonstrates the significance of the thermal decoupling of the magnetic moment and easy axis. 
The torques obtained from the MC sampling agree with the LLG results within the statistical uncertainty.
We then apply the hybrid model to the calculation of coupled spin and rotational dynamics. We finally study the frequency dependence of easy-axis reorientation, showing that the hybrid kMC--MC  model is computationally efficient at low frequencies used in applications such as MNH and MPI, which are inaccessible to LLG-based models because of their requirement for small integration steps to resolve the precessional motion. 
\subsection{Equilibrium torque analysis: thermal decoupling} \label{sec:torque}
\begin{figure}[htb!]
    \centering
    \includegraphics[width=1.0\linewidth]{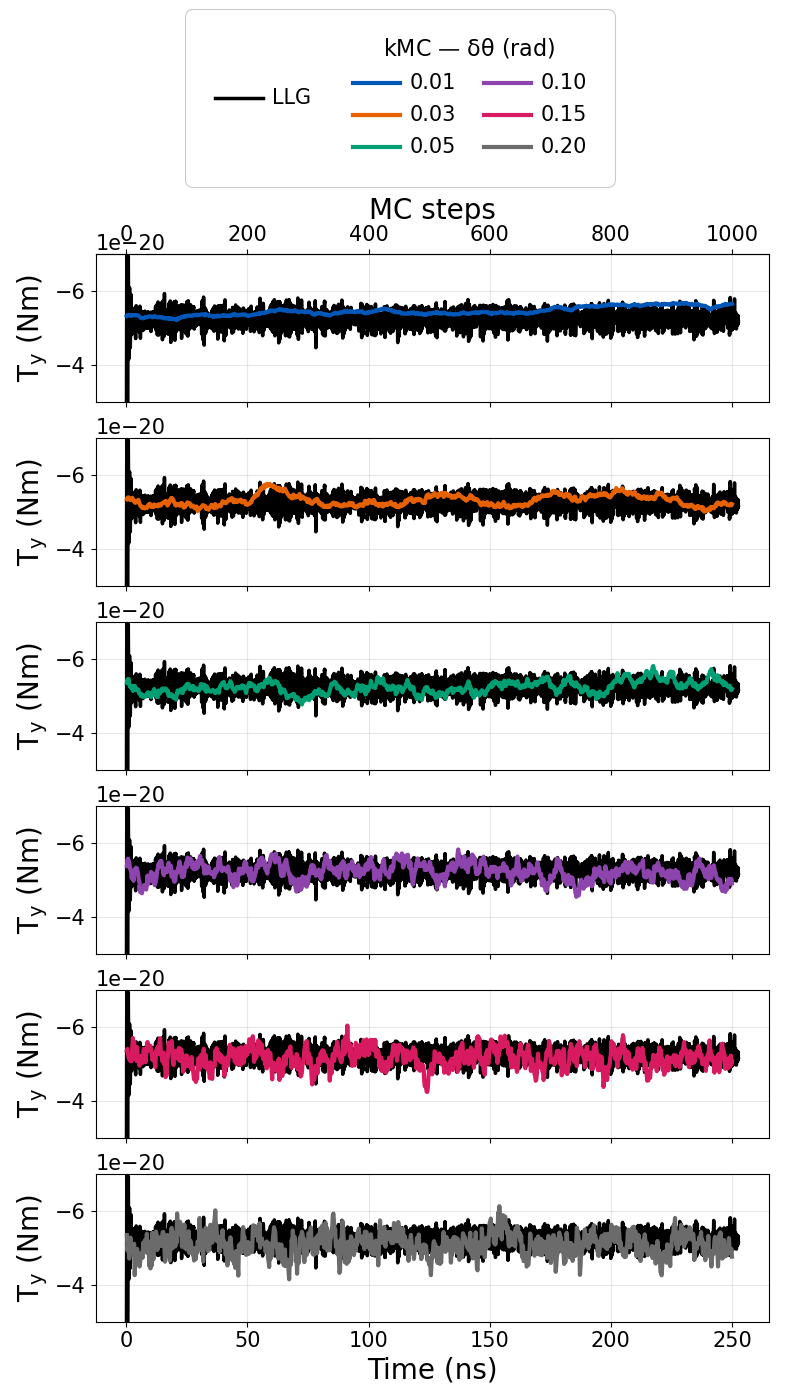}
    \caption{Time evolution of the anisotropy torque component, $T_y$, for 30~nm MNPs with the easy axis fixed at $\pi/6$~rad with respect to an applied field of $0.5 H_k$, with
    $KV/k_BT=37.5$. Each panel compares the stochastic LLG reference (black) with the hybrid kMC--MC result for a different MC cone angle, $\delta\theta = 0.01$--$0.20$~rad. The lower and upper horizontal axes show the real simulation time and the corresponding number of MC steps, respectively. 
    }
    \label{fig:torque_all}
\end{figure}

A critical parameter in our strategy for MC update of the magnetic moment is the cone angle $\delta \theta$, which controls the size of MC trial moves~\cite{Nowak2000TimeStep}. To validate the choice of cone angle and to see its effect on the simulation, we analyze whether the MC method correctly reproduces the anisotropy torque acting on MNPs across different cone angles. This validation isolates the magnetic moment trial algorithm from the rotational Brownian Dynamics, allowing us to test the MC strategy independently. In dynamical simulations using the stochastic LLG equation, the torque causes the magnetic moment to precess about the nearest energy minimum, thereby exploring the energy surface. The introduction of MC moves after each kMC step models this process and can be optimized by choosing suitable MC steps to reproduce the correct anisotropy torque $\boldsymbol{T}_\mathrm{ani}$. Therefore, it is necessary for the MC calculations to optimize the number of MC steps to reproduce the correct anisotropy torque $\boldsymbol{T}_\mathrm{ani}$ for physical consistency between the two methods. 
The rotational torque on the particle is averaged over the $N$ MC moves per Brownian time step as follows:
\begin{equation}
    \langle\boldsymbol{T}_{ani}\rangle=- 
    \frac{1}{N}
    \sum_{i=1}^N
    2KV\,(\mathbf{m}_i\cdot\mathbf{n}_i)\,
    (\mathbf{m}_i\times\mathbf{n}_i).
    \label{eq:torque_all}
\end{equation}
Note that the factor $\mathbf{m}_i\cdot\mathbf{n}_i$ in Eq.~\ref{eq:torque_all} reflects the fact that, due to thermal processes, the magnetic moment and easy axis are misaligned by an amount that depends on the anisotropy energy~\cite{Chantrell1983}, leading to a reduction of the rotational torque in a manner neglected in 'fixed-spin' models which assume that the magnetic moment is rigidly fixed to the easy axis. The Brownian Dynamics integrator responds to the thermally averaged torque, which can be significantly smaller than that calculated using a fixed-spin model.

To investigate the effects of the thermal decoupling of $\mathbf{m}$ and $\mathbf{n}$, we calculated $\langle\boldsymbol{T}_{ani}\rangle$ as a function of MC-steps for a particle with diameter of $d = 30$~nm and compared these results with the torque computed directly from the LLG equation under identical conditions. Figure~\ref{fig:torque_all} shows the  evolution of the torque component $\langle T_y \rangle$ at a constant field of $0.5H_k$ oriented at $\pi/6$~rad to the easy axis, revealing how torque equilibrates as a function of MC steps and cone angle. The results compare the hybrid BD--kMC method with the stochastic LLG results (black line) over 250 ns of simulation time, and 1000 MC steps, for different MC cone angles. 
The results demonstrate how the choice of cone angle influences the equilibration behavior, fluctuations and amplitude of the torque evolution. We observe that the anisotropy torque reaches equilibrium within a few MC steps. This indicates that the hybrid kMC--MC moves efficiently sample the magnetic phase space. 
The equilibrium torque obtained from the kMC approach agrees with the LLG result within the statistical uncertainty throughout the simulation, confirming the physical consistency of the method. For small cone angles (0.01–0.05 rad), the torque evolution remains smooth with minimal fluctuations due to the high acceptance rate of small magnetic moment updates. As the cone angle increases to intermediate values around 0.10–0.15 rad, 
the kMC and LLG calculations yield comparable equilibrium torque amplitudes, indicating that the equilibrium values are accurately reproduced. Increasing the cone angle further to $0.20$ rad introduces significant noise in the instantaneous torque due to larger trial displacements of the magnetic moment, although the average torque still remains consistent with the LLG results. Overall, smaller cone angles provide smoother trajectories, while larger cone angles improve sampling efficiency. 

\begin{figure}[htb!]
    \centering
    \includegraphics[width=1.0\linewidth]{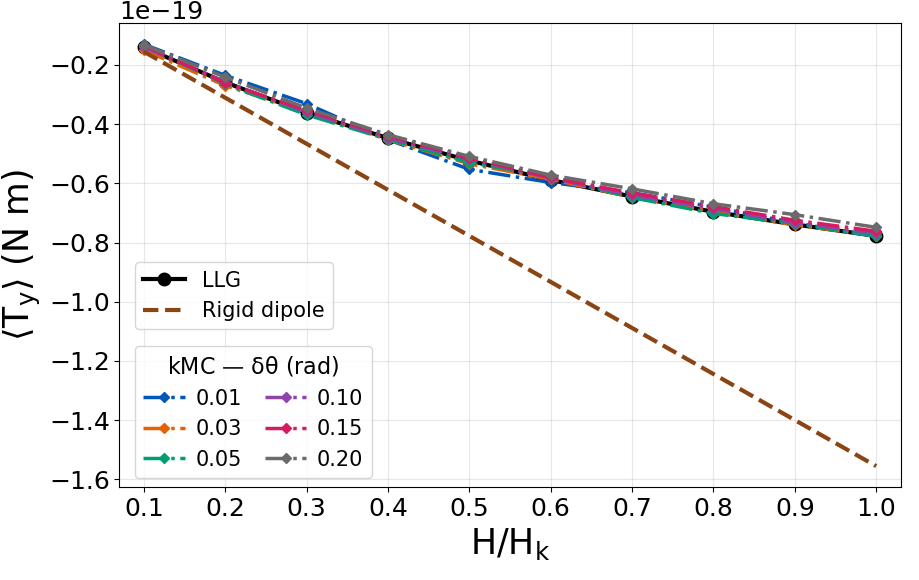}
   \caption{Average anisotropy torque component, $\langle T_y\rangle$, as a function of the normalized field amplitude $H/H_k$ for 30~nm MNPs with the easy axis fixed at $\pi/6$~rad with respect to the applied field. The solid black curve shows the stochastic LLG reference, while the colored curves correspond to the hybrid kMC--MC results for different MC cone angles, $\delta\theta = 0.01$--$0.20$~rad. The dashed line indicates the rigid-dipole reference, $T_y^{\mathrm{RD}}=-\mu_0M_sVH\sin\pi/6$, for the field and easy axis lying in the $xz$ plane.}
    \label{fig:torque_all_Hk}
\end{figure}

Figure~\ref{fig:torque_all_Hk} shows the dependence of the average anisotropy torque on the applied magnetic field strength for a field oriented at $\pi/6$~rad to the easy axis for different MC cone angles. For the kMC--MC simulations, the torque is averaged over the 1000 MC moves, whereas for the LLG simulations the average is calculated over the first 250~ns.
This allows us to directly examine how the equilibrium torque responds to the increasing field amplitude under identical sampling conditions. The results follow the same trend as observed in Fig.~\ref{fig:torque_all}:
for the larger cone angles considered, the average torque follows the LLG result (black solid line) over the investigated field range. 
This demonstrates that a unique range of cone angles (0.10–0.15) rad  can be used for field-dependent studies, ensuring consistent performance of the method under a varying external field. For comparison, the dashed line shows the rigid-dipole torque, $\mathbf{T}^{\mathrm{RD}}=\mu_0M_sV\,\mathbf{n}\times\mathbf{H}$, assuming that the magnetic moment is locked parallel to the easy axis. The smaller magnitude of the calculated average torque reflects the departure from rigid coupling.

Overall, the equilibrium torque analysis gives a clear picture. 
The thermally averaged anisotropy torque is consistent with the LLG result for all cone angles considered.
Because $\delta\theta$=0.10--0.15 rad reproduces the LLG equilibrium torque while maintaining efficient sampling, $\delta\theta=0.15$ rad is used in the subsequent simulations.
In addition, the torque is observed to reach equilibrium within less than 50 MC steps, suggesting rapid convergence of the sampling procedure. Therefore, a minimum of 100 MC steps is used to ensure proper averaging and statistical stability. An insufficient number of MC trial steps can lead to significant deviations from stochastic LLG results, as reported previously~\cite{okada2025proposal}. 
In summary, the results show that achieving the required kMC–LLG agreement necessitates an optimal balance between the cone angle, the sampling of thermal fluctuations, and the field-driven dynamics. 

This confirms that the method is efficient in describing relaxation within the local  minimum. However, the torque convergence alone is not sufficient to ensure the correct sampling of the equilibrium distribution. To address this limitation, we now perform a stricter validation under zero-field conditions, where the system is governed purely by anisotropy energy  and thermal fluctuations. The sampling can be directly tested against the analytical Boltzmann distribution  as a function of the reduced energy barrier $KV/k_B T$. 

\begin{figure}[htb!]
    \centering
    \includegraphics[width=1.0\linewidth]{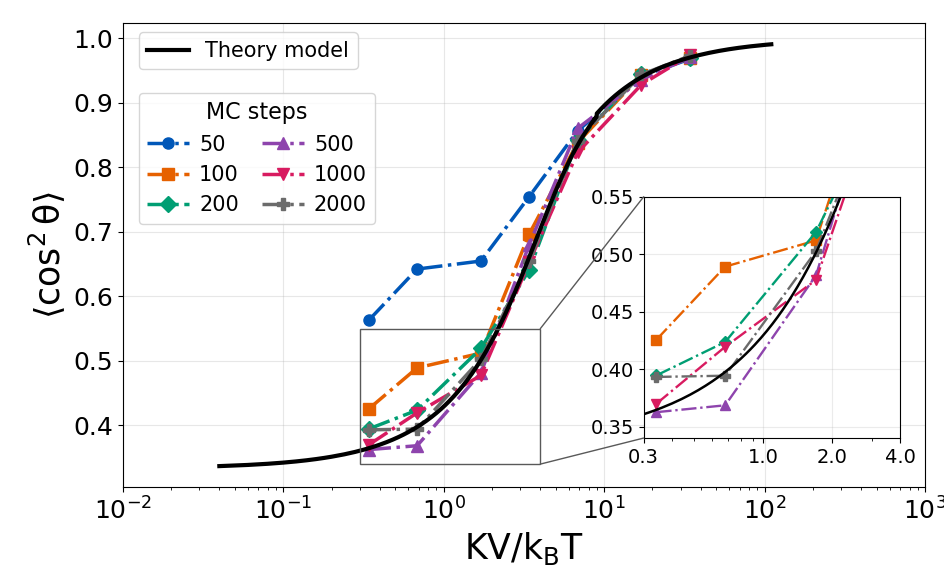}
    \caption{Orientational order parameter, $\langle \cos^{2}\theta \rangle$, plotted as a function of the reduced anisotropy barrier, $KV/k_B T$ in the absence of a magnetic field. The solid black curve denotes the theoretical prediction Eq.~\ref{equ:cos}, whereas the colored symbols represent the kMC simulation results using different numbers of MC steps. The comparison demonstrates that increasing the number of MC steps systematically reduces the deviation from the theoretical model, particularly in the low anisotropy barrier regime.}
    \label{fig:Roy-model}
\end{figure}

Figure~\ref{fig:Roy-model} shows the orientational order parameter $\langle \cos^2 \theta \rangle$, where \(\theta\) is the angle between the magnetic moment and the easy axis, as a function of $KV/k_B T$ for different MC steps, compared with the value calculated from the Boltzmann distribution, which is \cite{Chantrell1983}:
\begin{equation} \label{equ:cos}
   \langle\cos^2\theta\rangle =  \frac{1}{2\sqrt{\sigma }F(\sqrt{\sigma})} -\frac{1}{2\sigma},
\end{equation}
with $\sigma = KV/k_BT$ and the Dawson function
\begin{equation}
    F(x) = e^{-x^2} \int_0^x e^{t^2}dt.
    \label{equ:Dawson}
\end{equation}
The cone angle is fixed at 0.15 rad, as identified from the torque analysis. The results show that for small numbers of MC steps (50--100), significant deviations from the Boltzmann distribution appear, particularly in the intermediate barrier regime $KV/k_B T \sim 1\rightarrow 10$. In this region the thermal fluctuations dominate and accurate sampling within minima is essential. Although torque equilibrium is achieved in that range of MC steps, the sampling remains insufficient to reproduce the correct statistical distribution. However, this region corresponds to vanishingly small torque values. As the number of MC steps increases, the agreement with the theoretical prediction improves systematically. For 1000 and 2000 MC steps, the calculated order parameter approaches the Boltzmann prediction over the investigated range of $KV/k_B T$.
Using 2000 MC steps also ensures that in the case of small $KV/k_BT$ values, unphysically large fluctuations in the calculated torque will be eliminated.

This analysis demonstrates that torque convergence ensures fast local relaxation, while convergence to the Boltzmann distribution ensures correct equilibrium statistics. 
Based on the torque and Boltzmann distribution tests, $2000$ MC steps and a cone angle of $0.15$ rad were selected for the subsequent simulations.

\subsection{Dynamical behavior: coupling magnetic and Brownian degrees of freedom} \label{sec:brow}

Next, we extend the hybrid BD--kMC framework by incorporating the rotational Brownian Dynamics of the nanoparticle ensemble. The initial orientations of both the magnetic moments $\mathbf{m}_i$ and the easy axes $\mathbf{n}_i$ are randomly distributed over the unit sphere, while the applied AC magnetic field is directed along the $z$ axis. In particular, we analyze the time evolution of the averaged angle $\langle\theta_K\rangle$ between the easy axes $\mathbf{n}_i$ and the applied AC field $\mathbf{H}$ using different Brownian integration time steps:

\begin{equation}
    \langle\theta_K\rangle = 
    \frac{1}{N_p}\sum_{i=1}^{N_p}
    \arccos\!\left(
    \left|
    \mathbf{n}_i \cdot \hat{\mathbf{H}}
    \right|
    \right)
    \label{eq:avg_angle_def}
\end{equation}
where $\hat{\mathbf{H}}$ is the unit vector along the fixed direction of the applied AC field and $N_p$ is the total number of particles. The absolute value accounts for the
equivalence of $\mathbf{n}_i$ and $-\mathbf{n}_i$, mapping
angles above 90$^\circ$ onto their supplementary angles.
Thus, $\theta_K^f \in [0^\circ,90^\circ]$, with parallel and
antiparallel orientations treated equivalently.

The dynamics of $\langle\theta_K^f\rangle$ is governed by
two distinct mechanisms:
\begin{enumerate}
    \item \textbf{Thermally activated magnetic switching.} The magnetic moment evolves through thermally activated kMC switching events with transition rates given by Eq.~\ref{eq:switching}. The characteristic kMC timescale, $\tau_{\mathrm{kMC}}$, is defined as the inverse of the total transition rate. The attempt frequency is taken as $f_0 \sim 10^{9}\,\mathrm{s^{-1}}$, which represents the characteristic frequency of thermal activation and determines the average waiting time between successive switching events.
   \item \textbf{Brownian rotation of the nanoparticle.} The easy-axis orientation, $\mathbf{n}$, evolves through the combined effects of the anisotropy torque, $\boldsymbol{T}_{\mathrm{ani}}$, and thermal fluctuations. These torques induce rotational motion of the nanoparticle in the viscous carrier fluid. The characteristic rotational timescale is given by
    \begin{equation}
        \tau_{\mathrm{rot}} = \frac{\gamma_r}{6k_BT}
        = \frac{4\pi\eta r^3}{3k_BT},
        \label{eq:tau_rot}
    \end{equation}
    where $\gamma_r = 8\pi\eta r^3$ is the rotational friction coefficient, $\eta$ is the dynamic viscosity of the fluid, and $r$ is the particle radius.
\end{enumerate}
Here the relaxation time $\tau_\mathrm{rot}$ is found to be $\sim 3\, \mu s$, which is much larger than the characteristic timescales for magnetic switching, resulting in precise time integration of the rotational degrees of freedom.

\begin{figure}[htb!]
    \centering
    \includegraphics[width=1.0\linewidth]{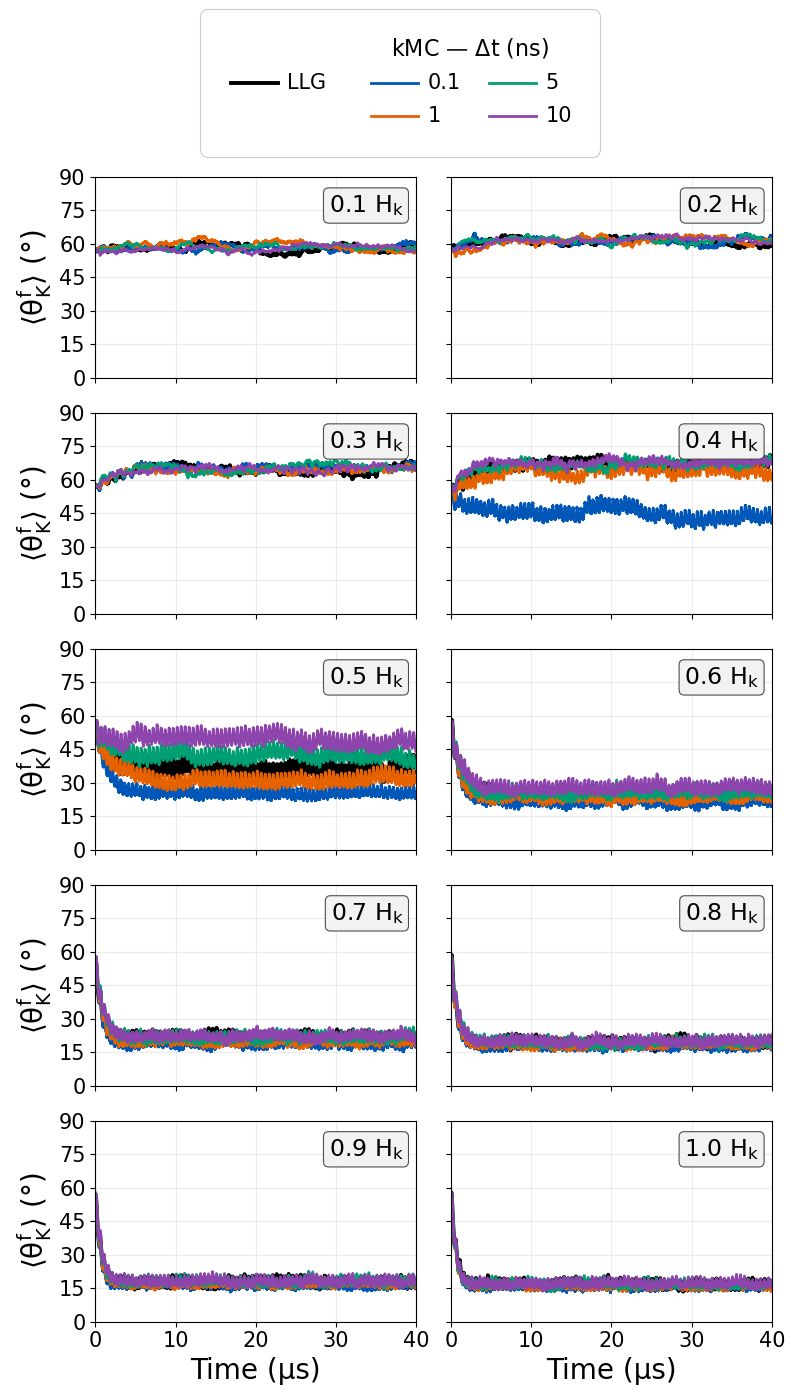}
    \caption{Time evolution of the  mean folded easy-axis angle, $\langle\theta_K^f\rangle$, for 30~nm MNPs under different normalized field amplitudes, $H/H_k = 0.1$--$1.0$, at a frequency of $f = 1~\mathrm{MHz}$. Each panel compares the stochastic LLG reference (black) with the hybrid kMC--MC results obtained using different Brownian integration time steps, $\Delta t = 0.1$, 1, 5, and 10~ns.}
    \label{fig:avg-angle-time-sizes}
\end{figure}

Fig.~\ref{fig:avg-angle-time-sizes} shows the time evolution of the mean folded easy-axis angle $\langle\theta_K^f\rangle$ calculated using the kMC--MC algorithm, compared with the LLG reference data (black solid line) for different applied field amplitudes. For all time steps $\langle\theta_K^f\rangle$ initially evolves rapidly before approaching a stationary oscillatory regime under the AC field. Around $H=0.5H_k$, however, there is a significant deviation from the LLG values and a strong, non-physical dependence on the chosen time step. 
To further investigate the calculated torque and its effect on the rotational motion, we calculate the stationary angle $\langle\theta_K^f\rangle_{\mathrm{final}}$ as a function of the normalized applied field. The results are shown in Supplementary Information S2, Fig. S2. The results based on the LLG calculations show, as in Ref~\cite{lopez2026ac}, that at low fields $\langle\theta_K^f\rangle_{\mathrm{final}}$ increases slowly, representing easy-axis orientation perpendicular to the applied field. With increasing field a transition toward an orientation parallel to the field is observed. In contrast, the hybrid BD--kMC model predicts different transition fields depending on the Brownian time step. In the following, we ascribe this effect to the MC steps leading to switching within the Brownian time step which is precluded in the LLG calculations by the slow evolution of the magnetization  within the minimum. This dynamical constraint is subsequently introduced approximately within the kMC--MC framework through the barrier-trapping approach.

\begin{figure}[htb!]
    \centering
    \includegraphics[width=1.0\linewidth]{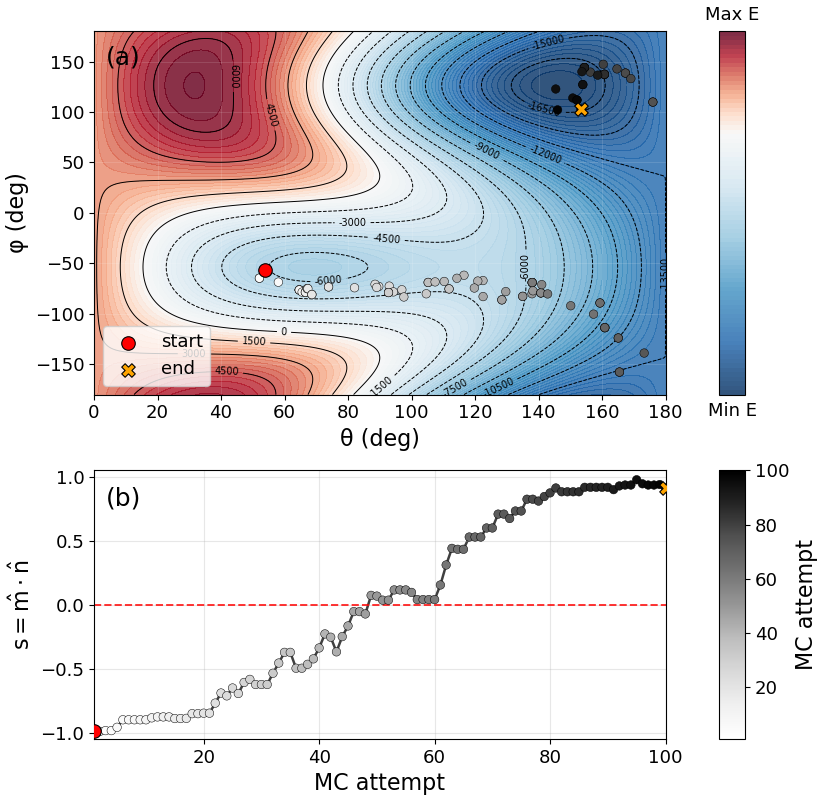}
    \caption{Illustration of a barrier-crossing event generated by Metropolis MC trial moves. Panel (a) shows the trajectory of the magnetic moment in the energy landscape during successive MC attempts, where the magnetic moment moves from one energy minimum to the opposite minimum. Panel (b) shows the corresponding evolution of $s=\mathbf{m}\cdot\mathbf{n}$ during the same sequence of MC attempts. For the specific particle and field configuration shown here, the applied field points along the negative $z$ direction and the easy axis forms an angle of $0.815$~rad with the field direction.}
    \label{fig:barrier-crossing-metropolis}
\end{figure}

\subsection{Switching within MC moves} \label{sec:swit}

As shown in Section S2 of the Supplementary Information, using conventional MC steps does not correctly couple the BD rotational dynamics with the kMC framework. We ascribe this to the timescales involved. While the MC moves correctly reproduce the equilibrium distribution of the magnetic moment directions, the LLG integration has insufficient time to reach the equilibrium, leading to an effective "dynamical constraint" which must be introduced into the MC process. In particular, a large number of switching events were found to occur during the MC trial moves, leading to significant deviations from the LLG prediction for the mean folded easy-axis angle, $\langle\theta_K^f\rangle$, as illustrated in the example shown in Fig.~\ref{fig:barrier-crossing-metropolis}. Physically, the thermally activated magnetic moment flipping across the barrier should be governed only by the kMC process, while the MC steps should describe only local thermal fluctuations within energy minima. Therefore, to estimate the correct $\langle\theta_K^f\rangle$ values, it is important to differentiate thermal fluctuations within the energy minima (MC steps) and magnetic moment flipping crossing the energy barrier (kMC rare events). If this separation is not maintained, barrier-crossing events can occur during the MC trial moves. As a result, unphysical switching events occur, leading to incorrect predictions of the rotational torque and consequently the average angle. To prevent this, in the following a barrier-trapping method is introduced in which a magnetic moment is restricted to its current local energy minimum during the MC steps.

\subsection{Barrier-Trapping Approach} \label{sec:barrier}

The barrier-trapping approach is implemented by introducing a criterion that confines the magnetic moment to its current energy well during the MC trial moves. In practice, this can be done by monitoring the orientation of magnetic moments with respect to the easy axis.  MC trial moves which cause the magnetic moment to cross the barrier should be rejected  so that only  fluctuations around the kMC energy minimum are allowed.  To implement this, we use  a simple condition based on the dot product of the easy axis and the magnetic moment $\mathbf{m} \cdot\mathbf{n}$. If the sign of this projection changes between the old and new configurations, the move is rejected. This criterion is illustrated in Fig.~S3 and can be expressed as
\begin{equation}
    \text{if}\quad
    (\mathbf{m}_\mathrm{old}\cdot\mathbf{n})\,
    (\mathbf{m}_\mathrm{new}\cdot\mathbf{n}) < 0 \quad
    \text{reject the attempt}.
    \label{eq:trap_condition}
\end{equation}
This condition ensures that the magnetic moment remains within the local energy minimum during the MC steps. The resulting barrier-trapping scheme allows the hybrid approach to reproduce the non-equilibrium easy-axis orientation obtained from the LLG simulations. 

\begin{figure}[htb!]
    \centering
    \includegraphics[width=1.0\linewidth]{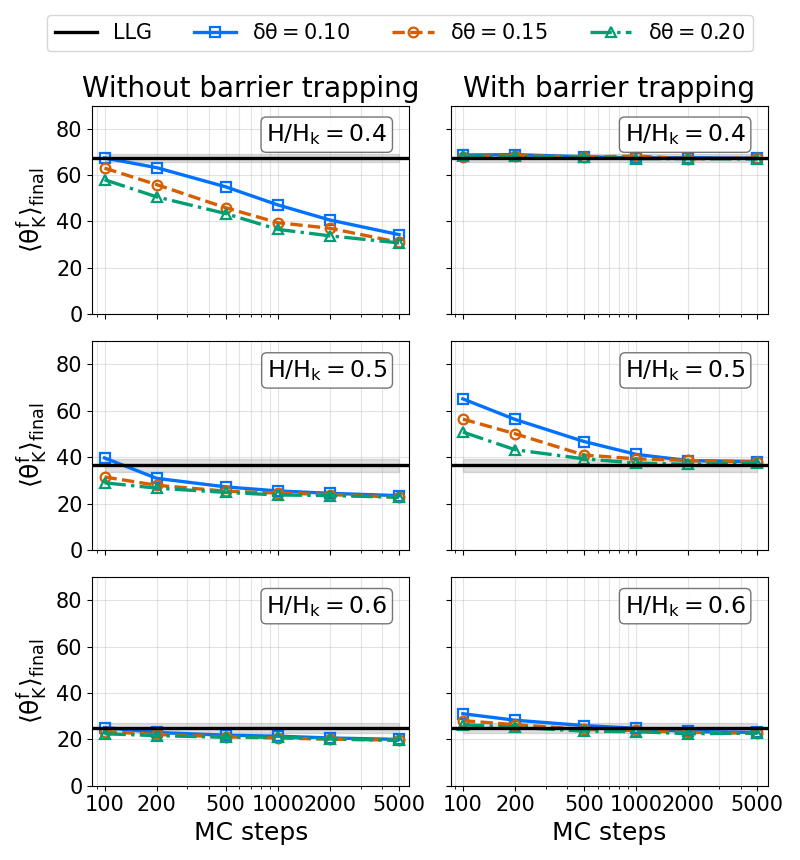}
    \caption{Stationary easy-axis orientation angle, $\langle \theta_K^f \rangle_{\mathrm{final}}$, as a function of the number of MC steps for 30~nm MNPs at a frequency of $f = 1~\mathrm{MHz}$. Results are shown for three normalized field amplitudes, $H/H_k = 0.4$, 0.5, and 0.6, and different MC cone angles, $\delta\theta = 0.10$, 0.15, and 0.20~rad. The left column shows the results obtained without barrier-trapping, whereas the right column shows the corresponding results with the barrier-trapping criterion applied. The solid black lines indicate the stochastic LLG reference values.}
    \label{fig:avg-angle-MCSTEPS-trap}
\end{figure}

Figure~\ref{fig:avg-angle-MCSTEPS-trap} shows the stationary easy-axis orientation angle  $\langle \theta_K^f \rangle_{\mathrm{final}}$ as a function of MC steps for different fields $H/H_k \approx 0.4, 0.5, 0.6$, using different cone angles (0.10, 0.15 and 0.20 rad). The stationary value is obtained by averaging the angle over the final 20~$\mu$s of the simulation, once the system has reached the stationary regime. The results are directly compared with LLG results in order to evaluate the hybrid BD--kMC model under different sampling conditions. Two different approaches are considered: conventional MC sampling without barrier-trapping and the modified scheme including barrier-trapping. Previously, it was observed in Fig.~\ref{fig:barrier-crossing-metropolis} that MC trial moves can occasionally cross the energy barrier directly during the MC updates. As a result, additional switching events are introduced independently of the kMC dynamics, leading to a deviation of $\langle \theta_K^f \rangle_{\mathrm{final}}$ from the results of LLG integration. This effect becomes larger with increasing cone angle and with increasing number of Metropolis MC steps. The deviation is particularly pronounced near the switching-field region in the range $H/H_k = 0.4$--$0.6$, where the barrier is reduced sufficiently that the system becomes more sensitive to thermal fluctuations. In all cases, the kMC--MC data deviate from the LLG data.
However, after implementing the barrier-trapping scheme, the calculated stationary angle converges to the LLG simulated values. All the results show convergence at 2000 MC steps and agree with the LLG data within the standard error.

\begin{figure}[htb!]
    \centering
    \includegraphics[width=1.0\linewidth]{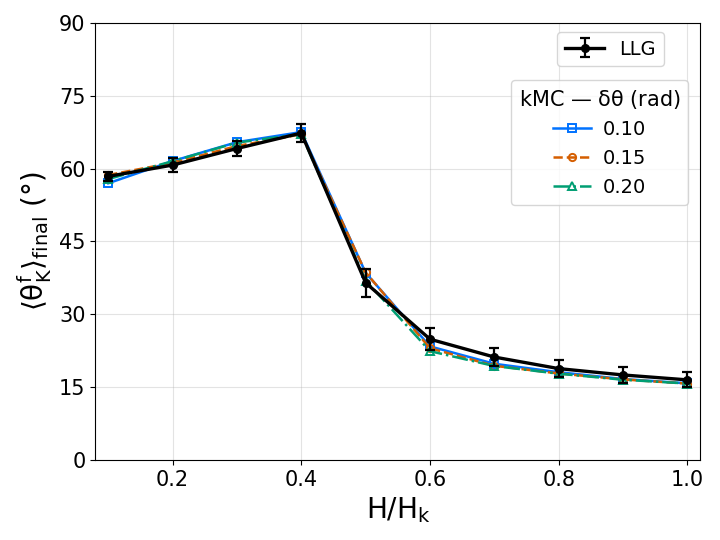}
    \caption{Stationary easy-axis orientation angle, $\langle \theta_K^f \rangle_{\mathrm{final}}$, as a function of the normalized field amplitude, $H/H_k$, for 30~nm MNPs at a frequency of $f = 1~\mathrm{MHz}$. The hybrid kMC--MC results are shown for different MC cone angles, $\delta\theta = 0.10$, 0.15, and 0.20~rad, using 2000 MC steps and the barrier-trapping criterion. The black curve shows the corresponding stochastic LLG results as a reference.}
    \label{fig:avg-angle-final-correct}
\end{figure}

Based on our analysis of different MC steps and cone angles in Fig.~\ref{fig:avg-angle-MCSTEPS-trap}, a cone angle of 0.15 rad and 2000 MC steps were selected for calculating the stationary angle $\langle\theta_K^f\rangle_{\mathrm{final}}$ as a function of $H/H_k$. Figure~\ref{fig:avg-angle-final-correct} presents $\langle \theta_K^f \rangle_{\mathrm{final}}$ after implementing the barrier-trapping scheme. With this correction, the previously observed mismatch between kMC and LLG in Fig.~\ref{fig:avg-angle-time-sizes} is now resolved. The barrier-trapping approach successfully restricts the thermal fluctuations within energy minima during MC steps, ensuring that the true switching events occur through the kMC transition. 
As a result, with a Brownian time step of $1.0$ ns, the barrier-trapped kMC results are consistent with the LLG data over the investigated field range.

\subsection{Frequency dependence of easy-axis reorientation} \label{sec:freque}

\begin{figure}[htb!]
    \centering
    \includegraphics[width=1.0\linewidth]{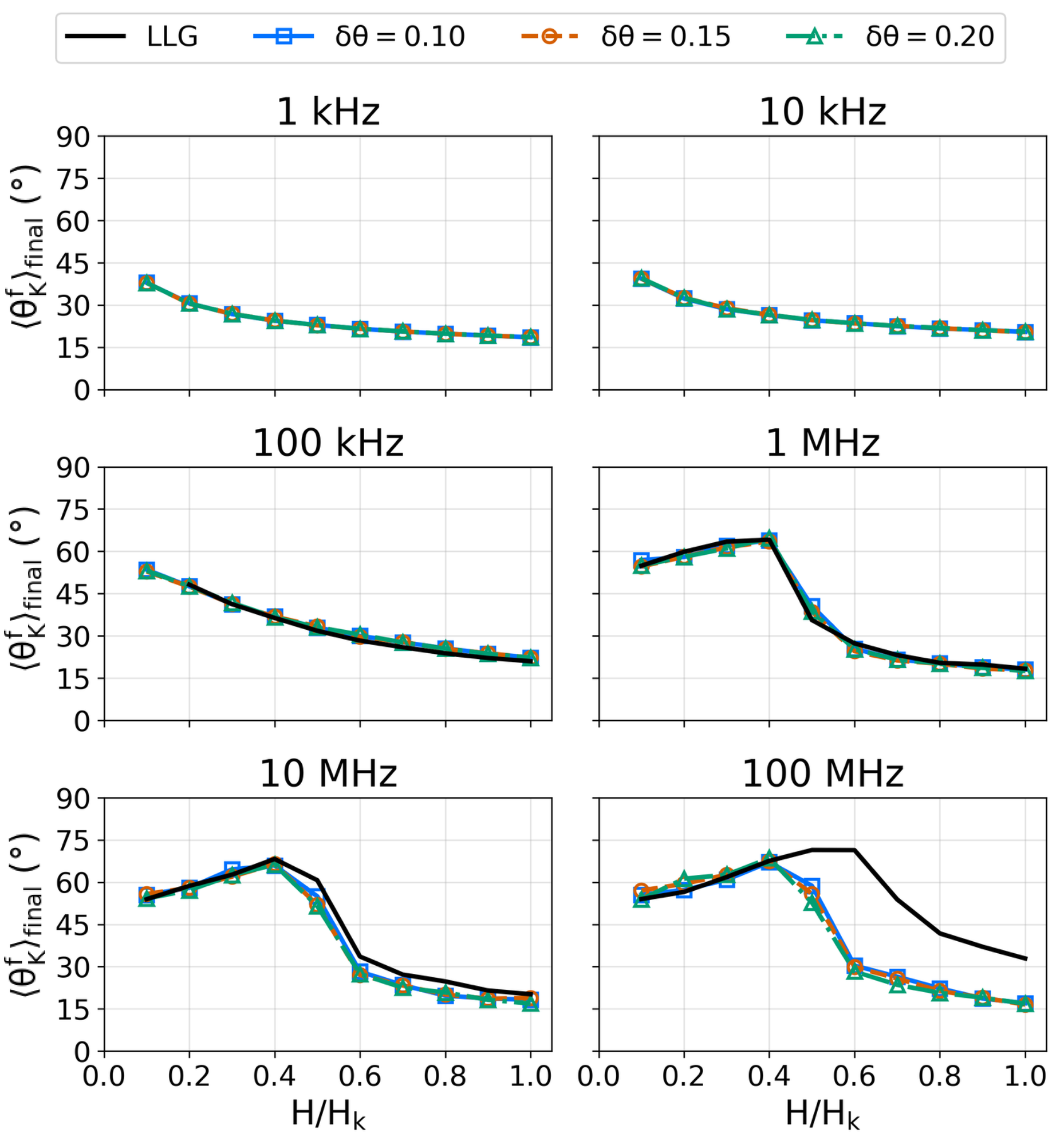}
    \caption{Stationary easy-axis orientation angle, $\langle \theta_K^f \rangle_{\mathrm{final}}$, as a function of the normalized field amplitude, $H/H_k$, for 30~nm MNPs. Each panel corresponds to a different excitation frequency, from $1~\mathrm{kHz}$ to $100~\mathrm{MHz}$. The colored curves show the hybrid kMC--MC results for different MC cone angles, $\delta\theta = 0.10$, 0.15, and 0.20~rad, obtained using 2000 MC steps and the barrier-trapping criterion. Black curves indicate the corresponding stochastic LLG reference results, where available.}
    \label{fig:avg-angle-final-freq}
\end{figure}

So far, our analysis has focused on an excitation frequency of
$f = 1~\mathrm{MHz}$, allowing us to systematically optimize the MC sampling parameters and validate the barrier-trapping approach against stochastic LLG simulations. We now extend the investigation to a broader frequency range, from $1~\mathrm{kHz}$ to $100~\mathrm{MHz}$, to examine the frequency dependence of easy-axis reorientation and assess the applicability of the hybrid kMC--MC approach under different dynamic conditions.

The appropriate method of integration depends on the frequency of the alternating field under investigation. Extremely high frequencies are expected to require the full precessional dynamics of the LLG integration approach, whereas low frequencies are more accessible to the kMC--MC dynamics due to increasing computational cost. Here we examine and compare the frequency-dependent stationary orientation in order to establish the optimal range for both techniques.

Figure~\ref{fig:avg-angle-final-freq} compares $\langle \theta_K^f \rangle_{\mathrm{final}}$ as a function of the normalized field amplitude $H/H_k$, for frequencies ranging from 1~kHz to 100~MHz along with different cone angles (0.10, 0.15, and 0.20~rad). This quantity characterizes the reorientation of the nanoparticle easy axis relative to the applied-field direction. 

Crucially, the field-induced transition between perpendicular and parallel orientation shown in previous figures is a key factor in all MNP applications.
 It also provides a crucial examination of models simulating the coupled spin and rotational dynamics.  At low-to-moderate frequencies (1~kHz--100~kHz), $\langle \theta_K^f \rangle_{\mathrm{final}}$ decreases monotonically with increasing $H/H_k$, consistent with the field progressively pulling the magnetization away from the easy axis toward closer alignment with the applied field direction. An important advantage of the kMC approach is its computational efficiency in this low-frequency regime. Because kMC directly models thermally activated barrier-crossing events rather than integrating the magnetization dynamics at every time step, it remains computationally tractable even for frequencies below 1~kHz, where the corresponding field periods become extremely long. In contrast, solving the stochastic LLG equation over such long physical times requires 
a number of integration steps that increases with the simulated field period, resulting in a high computational cost.
This significant reduction in computational cost enables kMC to investigate low-frequency magnetic phenomena that would otherwise be difficult to access using LLG. Such capability is particularly attractive for future studies of low-frequency magnetic applications, including MPI and other biomedical or sensing technologies where long-time magnetization dynamics plays an important role \cite{Harvell-Smith2022,Gavilan2025,Wilson2026,abbas2026simulation}.

A clear divergence between the kMC and LLG results is observed at 100 MHz. In the LLG simulations, $\langle \theta_K^f \rangle_{\mathrm{final}}$ broadens over a wide range of field amplitudes before gradually decreasing at larger $H/H_k$. 
In contrast, the kMC results deviate significantly from the LLG data. $\langle \theta_K^f \rangle_{\mathrm{final}}$ reaches its maximum at lower field amplitudes, subsequently decreasing at a faster rate than predicted by LLG results. 

This reflects the dominance of the precessional and damped magnetization dynamics in response to the rapidly oscillating magnetic field. Consequently, the switching field increases more rapidly than estimated by the kMC approach in this regime. 

This behavior arises because the kMC model represents magnetization reversal as discrete, thermally activated transitions between energy minima, omitting the continuous precessional motion and time-dependent dynamic relaxation described by the LLG equation. 
Consequently, the kMC approach reaches its limits of applicability in this regime.
Under these conditions, the magnetization dynamics are increasingly governed by deterministic precession and damping rather than purely thermally activated barrier crossing, thereby explaining the observed discrepancies.

\begin{figure}[htb!]
    \centering
    \includegraphics[width=1.0\linewidth]{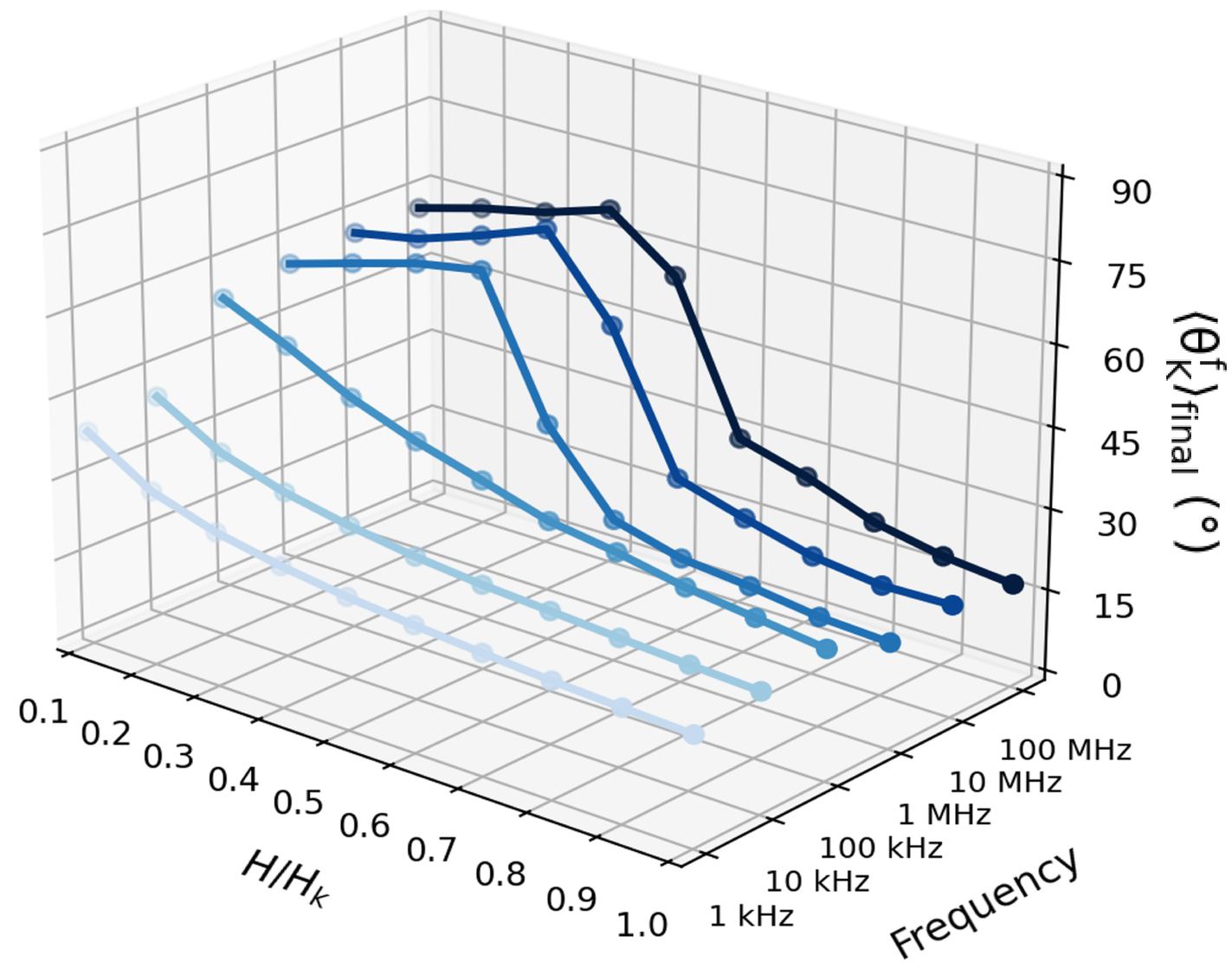}
    \caption{Three-dimensional representation of the stationary easy-axis orientation angle $\langle \theta_K^f \rangle_{\mathrm{final}}$ as a function of reduced field amplitude $H/H_k$ and frequency for an ensemble of 30 nm nanoparticles.}
    \label{fig:avg-angle-3d}
\end{figure}

Figure~\ref{fig:avg-angle-3d} provides a comprehensive three-dimensional representation of the stationary angle $\langle \theta_K^f \rangle_{\mathrm{final}}$. Unlike the two-dimensional plots presented in Fig.~\ref{fig:avg-angle-final-freq}, this figure illustrates the combined influence of field strength and frequency on the easy-axis reorientation dynamics of the nanoparticles. 
The plot clearly delineates 
the transition from a predominantly perpendicular easy-axis orientation, characterized by large values of
$\langle \theta_K^f \rangle_{\mathrm{final}}$, to a predominantly parallel configuration at lower angles.
As shown, this behavior remains consistent across all frequencies when \(H/H_k\) is below the switching threshold, where the applied field is insufficient to overcome the anisotropy energy barrier. 
In this regime, the anisotropy energy dominates the Zeeman energy, and the magnetic moment remains close to its initial equilibrium orientation~\cite{lopez2026ac}.

 As the field amplitude increases beyond the critical value, the nanoparticles undergo a rapid transition from perpendicular to parallel orientation, where $\langle \theta_K^f \rangle_{\mathrm{final}}$ drops sharply. This onset appears as a 
 shoulder in the curves, making it straightforward to identify where the reorientation occurs. 
 This threshold marks the critical field at which the Zeeman energy overcomes the anisotropy barrier, forcing the easy axis to align with the applied field.
 Another important feature is the shift of this switching field threshold with increasing frequency. At low frequencies, the transition from perpendicular to parallel orientation occurs at relatively small values of $H/H_k$. As the frequency increases, 
 the reorientation shifts toward higher field amplitudes, indicating that stronger fields are required to realign the particles.
Overall, the three-dimensional plot delineates the coupled evolution of the switching field and easy-axis reorientation as a function of frequency.

\section{Conclusions} \label{sec:conclusions}
We have developed a hybrid Brownian Dynamics--kinetic Monte Carlo (BD--kMC--MC) framework for simulating the coupled magnetic and rotational dynamics of non-interacting single-domain MNPs in a fluid environment under alternating  magnetic fields. The method combines Brownian Dynamics for particle rotation, kinetic Monte Carlo (kMC) for thermally activated magnetic switching, and local Metropolis MC sampling to account for thermal fluctuations of the magnetic moment within an energy minimum. This additional sampling allows the magnetization to fluctuate away from the easy axis, relaxing the fixed-spin constraint and accounting for the thermally averaged reduction of the anisotropy torque acting on the nanoparticle. The resulting thermally averaged torque is essential for consistently coupling Néel relaxation to Brownian rotation.

The MC sampling procedure was validated through equilibrium   anisotropy-torque calculations and comparisons with the analytical Boltzmann distribution. While the calculated torque equilibrates rapidly and agrees with the stochastic LLG results within statistical uncertainty, convergence of the torque alone does not guarantee correct equilibrium sampling. The combined analysis identified a cone angle of $\delta\theta=0.15$~rad and 2000 MC trial steps as suitable parameters for accurately sampling thermal fluctuations under the conditions investigated. These results emphasize the importance of intrawell magnetic fluctuations in determining the torque that couples magnetic relaxation to Brownian rotation.

Comparison with stochastic LLG simulations also revealed a limitation of unrestricted MC sampling in the coupled dynamics. 
Although MC moves reproduce the equilibrium distribution, they can generate barrier-crossing events that are already accounted for by the kMC transition rates. 
These additional switching events lead to incorrect predictions of the anisotropy torque and stationary easy-axis orientation, particularly near the switching-field region. 
We therefore introduced a barrier-trapping criterion that confines MC sampling to the current energy well, leaving interwell transitions to the kMC dynamics. 
This procedure effectively accounts for the dynamical constraint associated with the finite-time magnetic dynamics and restores agreement with the LLG results for the stationary easy-axis orientation over the investigated field range at 1~MHz.

Using the optimized model, we investigated the field- and frequency-dependent reorientation of the nanoparticle easy axes. At low field amplitudes, the easy axes preferentially orient perpendicular to the applied field, whereas increasing the field amplitude promotes a transition toward parallel orientations as Néel switching becomes increasingly important. The transition field also increases with excitation frequency, indicating that stronger fields are required to induce reorientation  at higher frequencies.
Comparison with stochastic LLG simulations shows that the hybrid approach reproduces the stationary easy-axis orientations over the investigated field range at frequencies up to approximately 10~MHz. 
At higher frequencies (100~MHz), however, the kMC-based model fails to reproduce the switching dynamics reliably, indicating the limitations of the rate-based description when the timescale of the excitation becomes incompatible with the assumptions underlying kMC.

The computational efficiency of the hybrid approach makes it particularly attractive for investigating coupled magnetic and Brownian Dynamics in the low-frequency regime, where direct LLG simulations become increasingly costly because of the need to resolve the magnetization dynamics over long physical times. Although a systematic comparison of computational efficiency is beyond the scope of the present study, an initial estimate   at 1~MHz indicates that the kMC--MC method is approximately 30 times faster than the corresponding LLG calculations under the computational conditions considered.

Overall, the results demonstrate that combining thermally activated switching with intrawell thermal fluctuations is essential for accurately describing the coupling between magnetic relaxation and Brownian rotation within the proposed framework. The resulting BD--kMC--MC approach provides a computationally efficient alternative to direct LLG simulations in the low- and intermediate-frequency regimes investigated, making it a promising tool for investigating MNP coupled magnetic and particle dynamics relevant to applications such as magnetic hyperthermia and magnetic particle imaging. 
Future extensions to interacting particles and polydisperse nanoparticle ensembles could further broaden its applicability.

\begin{acknowledgments}

This work was supported by the Spanish Ministry of Science, Innovation and Universities (MICIU) and the State Research Agency (AEI, 10.13039/501100011033) through Grants No. PID2024-157172NB-I00, No. PID2025-171111OB-I00, No. CNS2023-144321, and a Ramón y Cajal Fellowship (No. RYC2020-029822-I) to D.S. The Severo Ochoa Centres of Excellence program through Grant CEX2023-001286-S  funded by MICIU/AEI/10.13039/501100011033 is also acknowledged. Funding was also received by the Generalitat de Catalunya through Grant No. 2021SGR0032; by the Xunta de Galicia through Grants No. ED431F 2022/005 and No. ED431B 2023/055. This work was supported in part by JSPS KAKENHI Grant Number JP26K17308. We also acknowledge the Centro de Supercomputación de Galicia (CESGA) for providing computing resources.
\end{acknowledgments}

\bibliographystyle{apsrev4-2}
\bibliography{ref}             

\newpage
\onecolumngrid

\include{Supplementary_CondMat.tex}

\end{document}

%% file: Supplementary_CondMat.tex
\section*{Supplementary Information}

\renewcommand{\thefigure}{S\arabic{figure}}
\renewcommand{\thesection}{S\arabic{section}}
\setcounter{section}{0} 

\section{Schematic comparison of computational methods}
\begin{figure*}[htb!]
    \centering
    \includegraphics[width=0.9\textwidth]{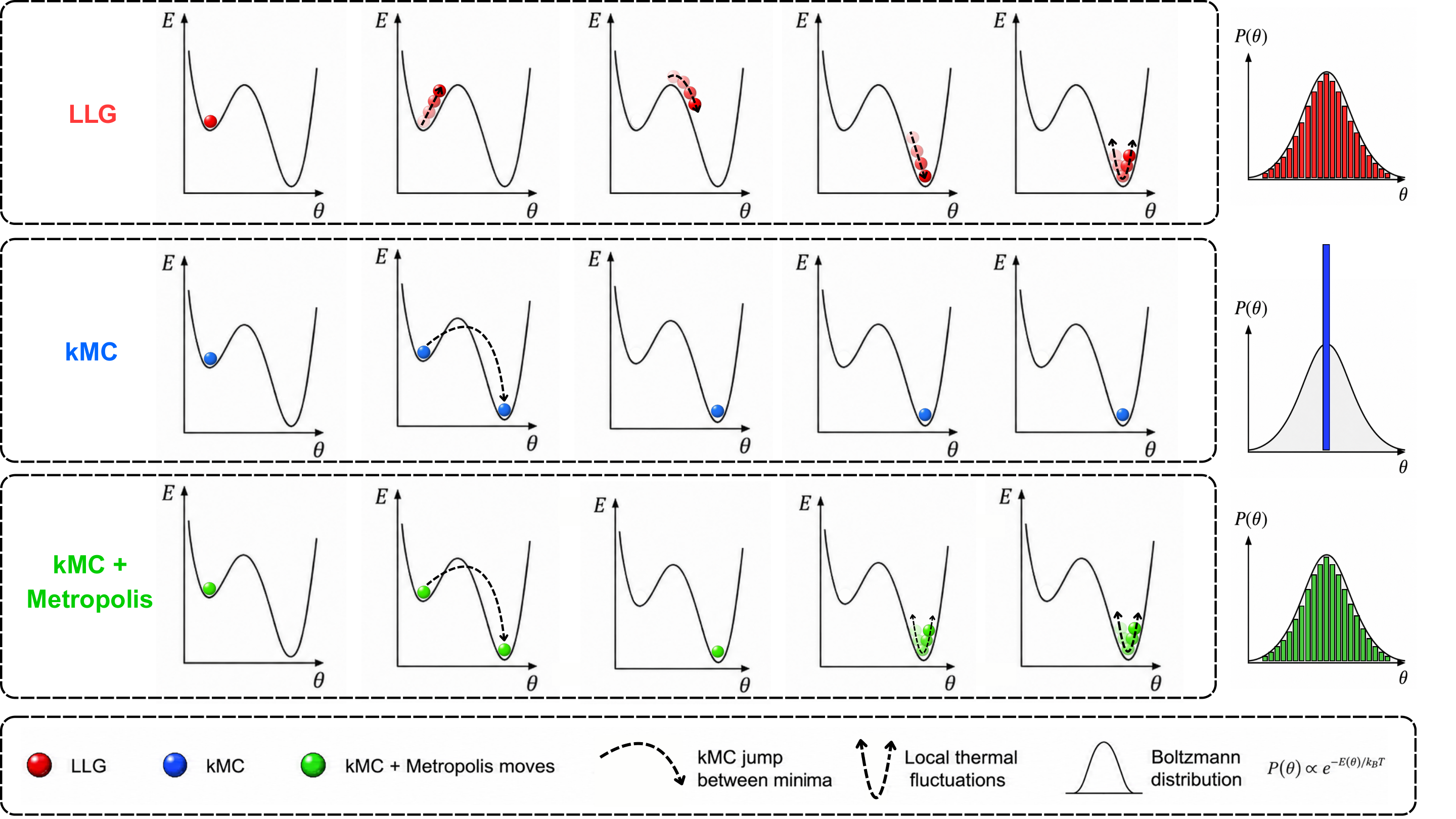}
    \caption{
    Schematic comparison of the magnetic moment evolution in the stochastic Landau--Lifshitz--Gilbert (LLG), kinetic Monte Carlo (kMC), and hybrid kMC--Metropolis approaches. The LLG method continuously samples the energy landscape and naturally reproduces the Boltzmann distribution through thermally driven fluctuations. In contrast, conventional kMC describes only thermally activated transitions between energy minima and therefore neglects local intra-well fluctuations. By incorporating Metropolis Monte Carlo moves into the kMC framework, local thermal fluctuations within each energy well are recovered, restoring the Boltzmann statistics and reproducing the effective stochastic behavior of LLG while maintaining significantly lower computational cost.
    }
    \label{fig:S1}
\end{figure*}

\section{Comparison of field dependent orientation using LLG and kMC-MC methods}

\begin{figure}[htb!]
    \centering
    \includegraphics[width=0.7\linewidth]{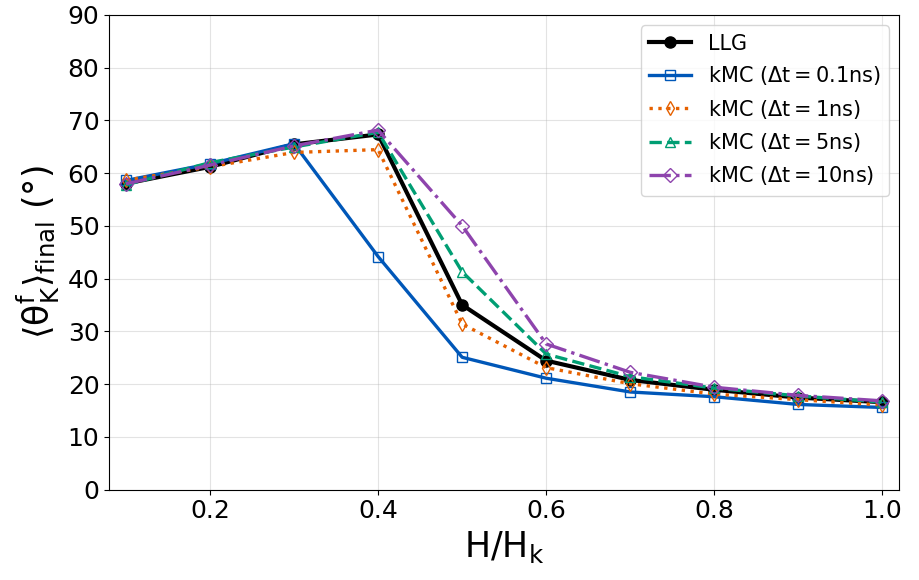}
    \caption{Average Angle orientation $\langle\theta_K^f\rangle_{\mathrm{final}}$ for particle size of 30 nm as function of normalized applied field ($0.1-1.0) H/H_k$. The different curves shows the kMC simulation data for using different Brownian timesteps, while the black curve represents the LLG data.}
    \label{fig:avg-angle-Hk-sizes}
\end{figure}

Stationary angle $\langle\theta_K^f\rangle_{\mathrm{final}}$ values  are plotted as function of normalized applied field in Figure~\ref{fig:avg-angle-Hk-sizes}. This representation allows to identify precisely the switching fields. Figure~\ref{fig:avg-angle-Hk-sizes} compares the results obtained using Brownian timesteps of (0.1 ns, 1 ns, 5 ns, and 10 ns) with the LLG reference (black solid line). The LLG data shows that at low fields ($H/H_k < 0.3$), the $\langle\theta_K^f\rangle_{\mathrm{final}}$ increases slowly, representing easy axis orientation perpendicular to the applied field. As the field strength increases, a peak appears at $H/H_k \approx 0.4$, representing the transition field where switching of the magnetization within that particles occurs. Beyond this transition region, the $\langle\theta_K^f\rangle_{\mathrm{final}}$ decreases sharply at high fields ($H/H_k \geq 0.6$), confirming the alignment of the easy axes with the applied field.

In contrast, The hybrid BD-kMC model predicts different transition fields depending on the Brownian timestep. For the smallest timestep, $\Delta t = 0.1$ ns (blue curve), the transition occurs at $H/H_k \approx 0.3$, which is lower than the LLG transition field of $H/H_k \approx 0.4$. Increasing the timestep to  $\Delta t = 1$ ns (orange curve) and $\Delta t = 5$ ns (green curve) shifts the transition fields to approximately $H/H_k \approx 0.4$, in close agreement with LLG, but there is still a slight deviation in transition region. While, the largest timestep, $\Delta t = 10$ ns (purple curve), overestimates the $\langle\theta_K^f\rangle_{\mathrm{final}}$ over the entire field range and fails to accurately reproduce the transition behavior.  The data for $\Delta t =0.1ns$ is included to show the effect of using timesteps smaller than the natural timescale of the kMC method, leading to non-physical results. We ascribe this effect to the MC steps leading to switching within the Brownian timestep which is precluded in the LLG calculations by the slow evolution of the magnetization density within the minimum. In the kMC method with MC moves it is shown that this 'dynamical constraint' can be approximately introduced by a 'barrier trapping' method. 


